\documentclass[a4paper,fleqn,usenatbib]{mnras}

\usepackage{mathptmx}
\usepackage[T1]{fontenc}
\usepackage{ae,aecompl}

\usepackage{graphicx}	% Including figure files
\usepackage{amsmath}	% Advanced maths commands
\usepackage{amssymb}	% Extra maths symbols
\usepackage{cite}
\usepackage{caption}
\usepackage{subfig}
\usepackage{mathtools}

\newcommand{\delphi}{$\Delta$\,--\,$\Phi$}

\title[The Eclipses of EX Draconis]{EX Draconis: Using Eclipses to Separate Outside-In and Inside-Out Outbursts}

\author[J.M.C. Court et al.]{
J.M.C. Court$^{1}$\thanks{E-mail: James.Court@ttu.edu},
S. Scaringi$^{1}$,
C. Littlefield$^{2}$,
N. Castro Segura$^{3}$,
K. S. Long$^{4}$,
\newauthor T. Maccarone$^{1}$,
D. Altamirano$^{3}$,
N. Degenaar$^{5}$,
R. Wijnands$^{5}$,
T. Shahbaz$^{6}$,
\newauthor Z. Zhan$^{7}$
\\\\
$^{1}$Department of Physics and Astronomy, Texas Tech University, PO Box 41051, Lubbock, TX 79409, USA\\
$^{2}$Department of Physics, University of Notre Dame, Notre Dame, IN 46556, USA\\
$^{3}$School of Physics and Astronomy, University of Southampton, Southampton SO17 1BJ, UK\\
$^{4}$Space Telescope Science Institute, 3700 San Martin Drive, Baltimore, MD 21218, USA\\
$^{5}$Anton Pannekoek Institute for Astronomy, University of Amsterdam, Postbus 94249, 1090 GE Amsterdam, The Netherlands\\
$^{6}$Instituto de Astrof\'{i}sica de Canarias (IAC), E-38205 La Laguna, Tenerife, Spain\\
$^{7}$Department of Earth, Atmospheric, and Planetary Sciences, M.I.T., Cambridge, MA 02139, USA\\
}

\date{Accepted XXX. Received YYY; in original form ZZZ}

\pubyear{2019}

\begin{document}
\label{firstpage}
\pagerange{\pageref{firstpage}--\pageref{lastpage}}
\maketitle

\begin{abstract}
We present a study of the eclipses in the accreting white dwarf EX Dra during \textit{TESS} Cycles 14 and 15.  During both of the two outbursts present in this dataset, the eclipses undergo a hysteretic loop in eclipse-depth/out-of-eclipse-flux space.  In each case, the direction in which the loops are executed strongly suggests an outburst which is triggered near the inner edge of the accretion disk and propagates outwards.  This in turn suggests that the outbursts in EX Dra are `Inside Out' outbursts; events predicted by previous hydrodynamic studies of dwarf nova accretion disks and confirmed spectroscopically in a number of other accreting white dwarf systems.  We therefore propose that the direction of the loop executed in eclipse-depth/out-of-eclipse flux space be used as a test to phenomenologically distinguish between `inside out' and `outside in' outbursts in other eclipsing dwarf novae; a reliable and purely photometric test to differentiate between these phenomena.
\end{abstract}


%%%%%%%%%%%%%%%%%%%%%%%%%%%%%%%%%%%%%%%%%%%%%%%%%%

%%%%%%%%%%%%%%%%% BODY OF PAPER %%%%%%%%%%%%%%%%%%

\section{Introduction}

\par EX Draconis \citep{Fiedler_EXDra} (hereafter EX Dra) is an Accreting White Dwarf (AWD) which is accreting matter from a Roche-lobe overflowing M dwarf star \citep{Knigge_Relation}.  Photometric studies have shown that the system has an orbital period of 5.04\,hr, and that the system is nearly edge-on with an inclination of $\sim85^\circ$ \citep{Baptista_EXDra}.  EX Dra regularly undergoes outbursts, with a recurrence time of $\sim30$ days, leading to this system being classified as a dwarf nova. %From AAVSO
\par The high inclination of EX Dra results in deep eclipses of the white dwarf, the accretion disk and the hotspot \citep{Joergens_EXDraComponents}, as well as shallower eclipses of the secondary star offset in orbital phase by 0.5 \citep[e.g.][]{Golysheva_LCAnalysis,Khruzina_EXDra}.  The large depths of these eclipses make EX Dra a good candidate on which to perform a study of eclipse variation over time, a technique which can shed light on the behaviour of the accretion disk \citep[e.g.][]{Smak_UGem} and help us better understand the nature of the outbursts in these objects.  Previous studies of EX Dra, such as the study by \citet{Billington_EXDra} shortly after the discovery of the source, have focused on the shape of the eclipses and used them to place constraints on the parameters of the two objects in the binary.  In particular these authors found that the flux at mid-eclipse during quiescence is consistent with being entirely from the companion star, suggesting that the accretion disk is entirely eclipsed at this time and hence the radius of the accretion disk is smaller than that of the donor star.
\par Dwarf nova outbursts such as those seen in EX Dra can be divided into two categories depending on the evolution of the accretion disk throughout the event:  so-called `Outside-In' outbursts, which begin partway through an accretion disk at a substantial distance from the white dwarf and propagate inwards, and `Inside-Out' outbursts, which begin near the disk's inner edge and propagate out \citep[e.g.][]{Meyer_Nova}.  Both types of outburst have been modelled mathematically by \citet{Mineshige_OIIO}.  They suggest that in an Outside-In outburst, there is first a build-up of matter at the outer edge of the disk due a low disk viscosity which does not allow for efficient outward transfer of angular momentum.  The surface density thus increases at this radius until no stable configuration in the `cold' state exists, at which point this part of the outer disk switches to the `hot' state (see e.g. \citealp{Ichimaru_DiskStates,Mineshige_S}).  This triggers a heating wave which propagates inwards, until it either dissipates or the entire disk is in the `hot' state.  While most AWDs preferentially show outbursts of one type or the other, a number of systems (e.g. NY Ser, \citealp{Sklyanov_NYSer}) have been observed to undergo both Inside-Out and Outside-In outbursts.
\par The models of \citet{Mineshige_OIIO} show that Inside-Out outbursts should occur at lower accretion rate $\dot{M}$ (e.g. \citealp{Mineshige_S}).  In this case, the mass transfer rate is low enough that material does not pile up at the edge of the disk and is able to move inwards via viscous diffusion, increasing surface density globally throughout the disk.  The critical surface density value above which no cold state solution exists is smaller in the inner disk, and hence the instability is first triggered relatively close to the inner edge of this disk.  From this point, the heating wave propagates both inwards and outwards until the entire disk is in the hot state.
\par Although the models of \citet{Mineshige_OIIO} provide physically motivated explanations for how each type of outburst occurs, they make a number of predictions which have not been supported by subsequent observations \citep{BuatMenard_NovaModelFix}.  If the type of outburst is determined only by the global accretion rate, for example, it would not be expected that individual systems would show both types of outburst without corresponding significant evolutionary changes.  Modifications to the models (e.g. \citealp{BuatMenard_NovaModelFix}) have so far been unable to explain the observed fact that both types of outburst seem to be able to occur in systems with both low and high accretion rate, meaning that the trigger criteria behind Inside-Out and Outside-In outbursts remains an open question.
\par Inside-Out and Outside-In outbursts can be distinguished with high-quality two-colour observations \citep[e.g.][]{Ioannou_OutsideIn,Webb_InsideOut}.  In the absence of multiple colours, however, the common way to separate the two types of outburst is with qualitative parameters such as the `shape' of each outburst.  This method of separation outburst types has been noted as being inaccurate, subjective and heavily affected by human biases \citep[e.g.][]{Kato_3DN}.  As such, a new model-independent photometric test to differentiate these two phenomena is required to build a robust sample of each type of outburst in order to perform future studies to better understand their physical differences.
\par It has been shown \citep[e.g.][]{Smak_UGem,Rutten_OYCar} that, in eclipsing AWDs, the profile of the eclipses can also be used to distinguish between Inside-Out and Outside-In outbursts; as the secondary star passes in front of the accretion disk, different parts of the disk are eclipsed in sequence, and hence analysing the shape of the eclipses can indicate which radii in the disk are brighter than others.  \citet{Baptista_EXDra} used this technique on a number of outbursts of EX Dra in 1995-1996 using a small sample ($\sim30$) of observed eclipses.  They found that the outbursts in EX Dra were all consistent with being Inside-Out in nature.  However, this method of determining outburst type requires detailed modelling of individual eclipses, which in turn relies on high-quality lightcurve data with time resolutions on the order of $\sim10$\,s, and hence the number of systems on which such analysis can be performed is low.
\par EX Draconis is among the objects currently being observed by \textit{TESS} during its survey of transients in the northern hemisphere.  Due to the high observation cadence and long stare-time of this instrument, we now have an unprecedented large sample of eclipses from this object, during both outbursts and quiescent periods, allowing population studies of these events as well as studies of how they evolve over time.  In this paper we present such a study of the eclipses in this object, and we show that a study of eclipses in a dwarf nova can be used to differentiate between the aforementioned Outside-In and Inside-Out outbursts in these systems without needing to assume complex eclipse profiles a priori.
\par AWDs can also be studied in the timing domain.  These systems naturally show a strong Fourier peak corresponding to their orbital period, but many AWDs also show evidence of a distinct signal at a slightly higher or lower frequency.  This modulation, referred to as a `superhump', is believed to be a beat frequency between the orbital period and some form of modulation in the accretion disk.
\par Superhumps come in two different varieties.  `Positive' superhumps, in which the period of the modulation is greater than the orbital period, are seen during unusually large `super'-outbursts of a class of dwarf nova systems referred to as SU UMa-like AWDs.  This type of superhump is believed to be caused by the nodal precession of an elongated accretion disk \citep[e.g.][]{Horne_Superhump}.  Negative superhumps on the other hand, in which the period of the modulation is shorter than the orbital period, are seen during quiescence and smaller outbursts in a wider variety of AWDs \citep[e.g.][]{Harvey_NSH}.  This type of superhump is generally interpreted as being caused by the vertical precession of a \textit{tilted} accretion disk \citep[e.g.][]{Bonnet-Bidaud_NegativeSuperhumps}.
\par The geometric origins of superhumps mean that their presence, or absence, in a system can provide valuable information as to the configuration of the accretion disk in that system.  As such we also probe the properties of the superhumps of EX Dra in this paper, giving us additional information on the geometry of the system and helping to explain how and why outbursts in this system evolve in the way that they do.

\section{Observations \& Data Analysis}

\subsection{Dataset}

\par We make use of data from the Transiting Exoplanet Survey Satellite (\textit{TESS}, \citealp{Ricker_TESS}); a space-based observatory launched in 2018.  \textit{TESS}'s primary mission is to perform a survey of optical transits in both the both the northern and southern ecliptic hemispheres.  It does this by dividing the sky into a number of sectors, each of which is observed for a period of $\sim1$ month.  The entire sector is observed at a cadence of 30 minutes and a number of pre-selected objects, including EX Dra, are each observed at a cadence of 2 minutes.  Both the short-cadence and long-cadence barycentred data are freely available from the portal provided by the Mikulski Archive for Space Telescopes (MAST\footnote{\url{https://mast.stsci.edu/portal/Mashup/Clients/Mast/Portal.html}}).
\par
\begin{figure*}
    \includegraphics[width=2.2\columnwidth, trim = 0mm 0mm 0mm 0mm]{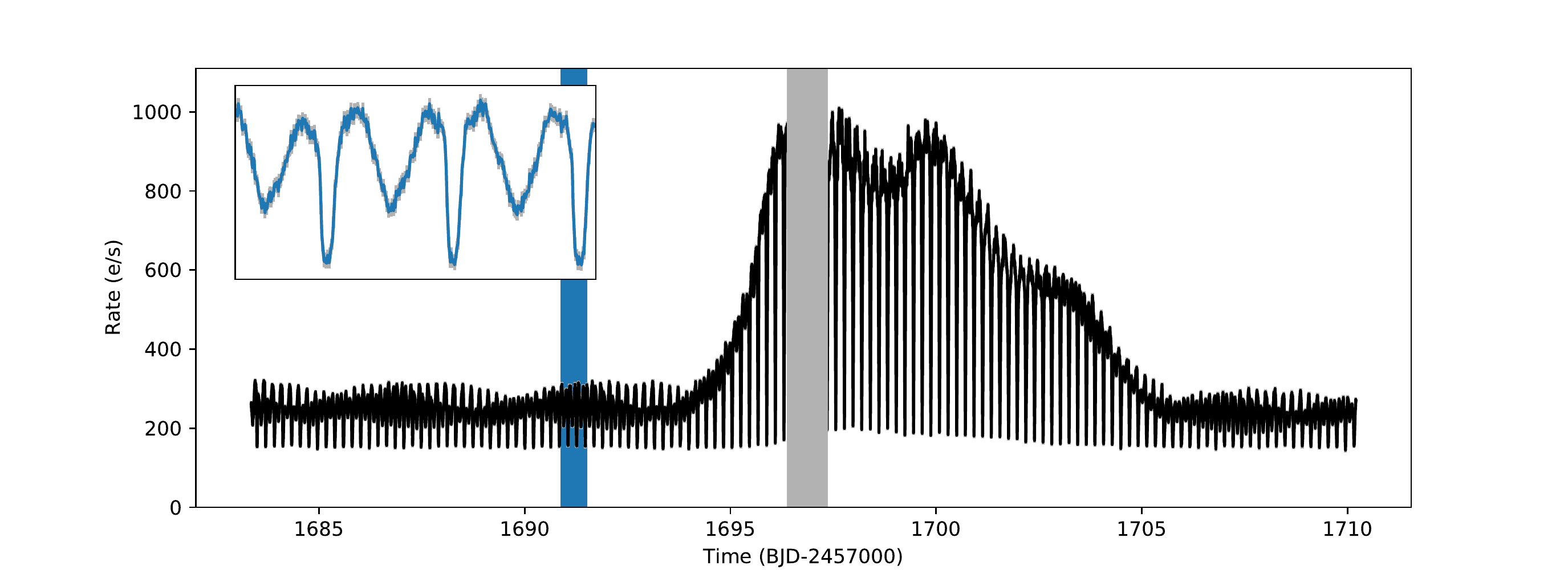}
    \includegraphics[width=2.2\columnwidth, trim = 0mm 0mm 0mm 0mm]{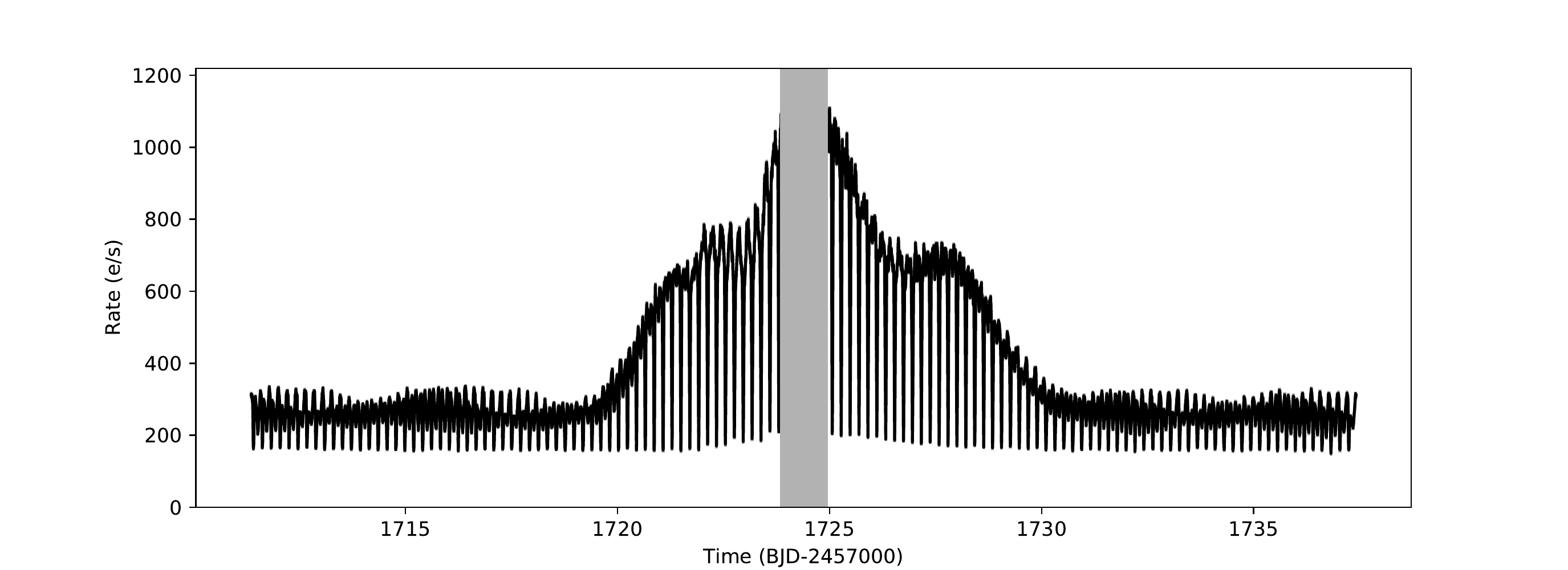}
    \captionsetup{singlelinecheck=off}
    \caption{Lightcurves of EX Dra from \textit{TESS} Sectors 14 (top) and 15 (bottom), showing one outburst in each.  Inset in the lightcurve of Sector 14 is a zoom to the region highlighted in blue, showing the structure of the eclipses in this source.  Grey periods indicate data gaps caused by telemetry constraints.  BJD refers to Barycentred Julian Days.}
   \label{fig:lightcurves}
\end{figure*}

 EX Dra was observed by \textit{TESS} during Sectors 14 \& 15, corresponding to the first two sectors in the survey of the northern ecliptic hemisphere; these in turn correspond to MJDs (Modified Julian Dates) 58682--58709 and 58710--58736 for a total of 53 days.  We show the lightcurves of these observations in Figure \ref{fig:lightcurves}.  While it was being observed, EX Dra underwent two outbursts; once in Sector 14, beginning at MJD $\sim58693$, and once in Sector 15, beginning at MJD $\sim58718$.  In both cases the beginning and the end of the outburst were observed, but a portion of the middle of the outburst fell into a $\sim1$ day data gap caused by \textit{TESS} telemetering data back to Earth.

\subsection{Eclipse Analysis}
\par The observations with \textit{TESS} in these two sectors include 239 eclipses: 122 in Sector 14 and 117 in Sector 15.  In order to analyse the properties of the eclipses, we first isolated them using our own algorithm:
\label{sec:algo}
\begin{enumerate}
\item Create a `smoothed' lightcurve by removing all oscillations at the orbital period:
\begin{enumerate}
\item For each datapoint $f(t)$ in the original lightcurve, make a subset of all datapoints $f(x)$ such that $t-\frac{p}{2}\leq x<t+\frac{p}{2}$, where $p$ is the orbital period of EX Dra.
\item Calculate the $n^{\textrm{th}}$ percentile rate value in $f(x)$, where $n$ corresponds to the approximate percentage of each orbital period in which the white dwarf and accretion disk are at least partly eclipsed.  Call this value $q_l$  For EX Dra, we use $n=20$.
\item Remove all datapoints with flux values less than or equal to $q_l$.  As the main eclipse is always the faintest part of each orbital period, this essentially removes the eclipse from the dataset, as well as the secondary eclipse if present.  This in turn prevents a change in the depth of an eclipse from artificially causing a change in the out-of-eclipse flux value measured in Step (iii).
\item Find the mean of the remaining datapoints in $f(x)$.  Replace the flux value of $f(t)$ with this value.
\end{enumerate}

\item Subtract the smoothed lightcurve from the original lightcurve to obtain the detrended lightcurve, or a lightcurve which only retains variability at the orbital period of the system.
\item For each eclipse in the detrended lightcurve, select a small time window around the eclipse minimum, such that the shape of lightcurve within does not include the ingress or egress of the hotspot.  Fit a Gaussian to the lightcurve in this period to extract values for eclipse depth and width.  Using the same small time range, make a subset of data from the smoothed lightcurve and take the mean flux of this subset to be the out-of-eclipse flux.
\end{enumerate}

\begin{figure}
    \includegraphics[width=0.9\columnwidth, trim = 3mm 0mm 16mm 10mm,clip]{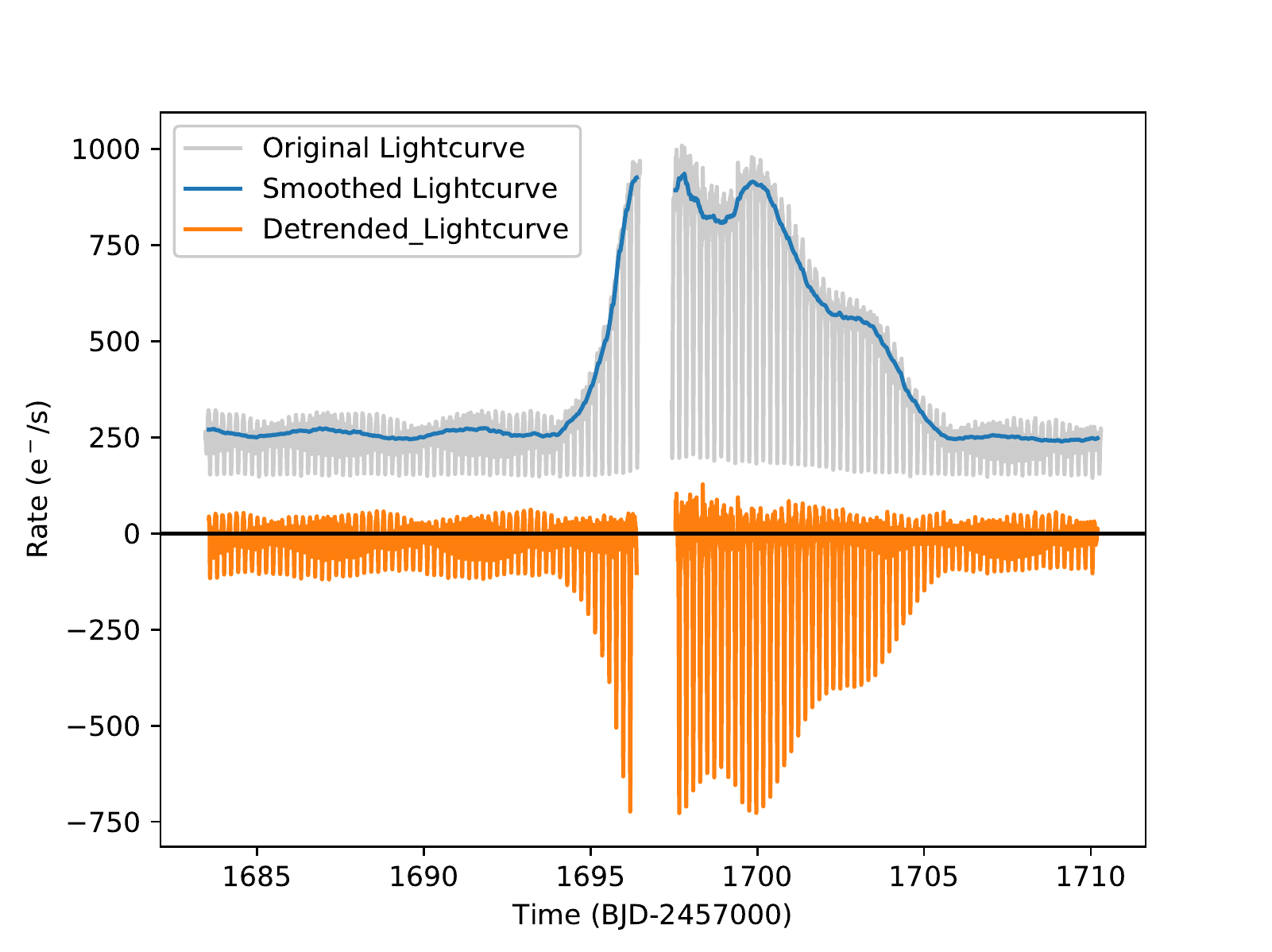}
    \captionsetup{singlelinecheck=off}
    \caption{The decomposition of the lightcurve of Sector 14 into smoothed and detrended components, using the algorithm outlined in Section \ref{sec:algo}.}
   \label{fig:detrending}
\end{figure}

In Figure \ref{fig:detrending} we show how our algorithm decomposes the lightcurve from Sector 14 into smoothed and detrended components.  Note that the eclipse-isolating algorithm we use here is different to the method used in \citet{Court_ZCha} to isolate eclipses in the AWD Z Cha.  In that paper, the authors remove the eclipses from the lightcurve and fit splines across the resulting data gaps to interpolate what the flux would be at each eclipse midpoint if no eclipse occured.  This method was valid in in Z Cha because, aside from a weak and relatively constant-brightness hotspot during quiescence, there was no significant modulation in the lightcurve on the orbital period.  By using splines to estimate the uneclipsed flux during data gaps caused by the removal of eclipses, the authors were able to account for variability of the source flux over timescales longer than an orbital period and hence more accurately estimate what the uneclipsed fluxes were at these times.  However in EX Dra the hotspot varies significantly during quiescence (e.g. \citealp{Golysheva_LCAnalysis}), as does the depth of the secondary eclipse and the brightness of the flux between the main and secondary eclipses, which would significantly contaminate our estimates for out-of-eclipse flux for each eclipse if we used the method of \citet{Court_ZCha}.

\section{Results}

\subsection{Period and Superhumps}
\label{sec:shump}
\par We calculate an independent value for the orbital period of EX Dra by first estimating a period from the largest peak in a Lomb-Scargle spectrum of the entire dataset from both observations.  We then folded the lightcurve over a range of periods close to this estimate, choosing the period which gave the lightcurve with the lowest dispersion.  We iterated this process for successively smaller ranges of periods until the change in dispersion between periods was no longer significant.  Using this method we found a period of 0.2099385(6) days, or 5.03852(1) hours.  This is similar to but slightly longer than previous periods reported for this object (e.g. 0.20993698(1)\,d, \citealp{Baptista_EXDra}).  This discrepancy is consistent with previously reported variations in the period of this system, including sinusoidal variations on timescales of years, and a long term trend towards higher ortbital period \citep{Pilarcik_EXDra,Baptista_EXDra}.
\par
\begin{figure}
    \includegraphics[width=\columnwidth, trim = 2mm 3mm 12mm 8mm]{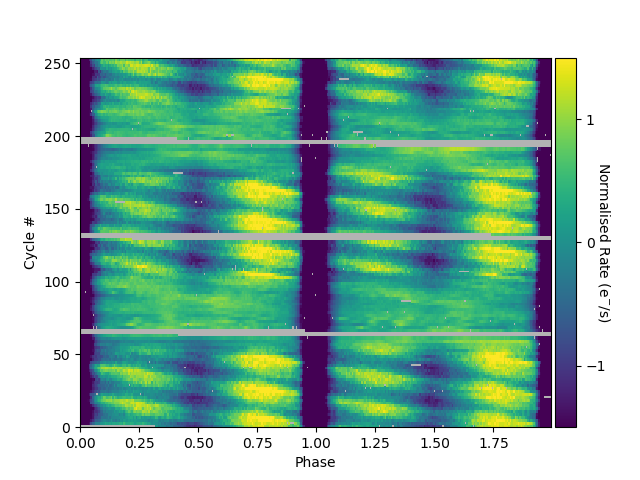}
    \captionsetup{singlelinecheck=off}
    \caption{A flux-phase diagram showing the \textit{TESS} lightcurve of EX Dra folded over a period of 0.2099385\,d, corresponding to the orbital period of the system.  Eclipses can be seen as dark vertical tracks on this plot centred at phase 0, whereas the superhumps can be seen as bright diagonal stripes.  The count rate in each orbital period has been normalised to better show the periodic behaviour of the source rather than longer-term variability.}
   \label{fig:flux_phase}
\end{figure}

 The \textit{TESS} lightcurves of EX Dra show evidence of a strong negative superhump in this source during quiescence with a period of $\sim4.81$\,hr.  These appear as diagonal `lines' in a flux-phase plot (shown in Figure \ref{fig:flux_phase}), which indicate a periodic signal offset slightly in frequency from the orbital frequency.  We find no evidence of either positive or negative superhumps during either outburst.

\par In order to test whether the superhump frequency changes during the observations covered in this study, we used generalised Lomb-Scargle spectroscopy \citep{Lomb_LombScargle,Scargle_LombScargle,Irwin_LombScargle} to calculate superhump periods separately from three different segments of the lightcurve:
\begin{itemize}
\item Segment 1 between the start of the lightcurve and MJD 2458694, corresponding to the quiescent period before the first outburst.
\item Segment 2 between MJDs 2458706 and 2458719, corresponding to the quiescent period between the two outbursts.
\item Segment 3 between MJD 2458731 and the end of the lightcurve, corresponding to the quiescent period after the second outburst.
\end{itemize}
\par In the Lomb-Scargle spectrum of each lightcurve segment, we fit a Gaussian to the peak corresponding to the superhump frequency.  We find peak superhump frequencies of 4.9901(8), 4.9900(2) and 5.014(6)\,BJD$^{-1}$ for Segments 1, 2 and 3 respectively\footnote{BJDs: Barycentred Julian Days}.  The superhump frequencies calculated for Segments 1 and 2 are consistent with being identical.  However, the frequency in Segment 3 is significantly higher.  The Gaussian fit to the superhump frequency in Segment 3 is also significantly broader, with a Gaussian width of 6.369(9)\,BJD$^{-1}$, compared to widths of 3.1541(7) and 2.9041(6)\,BJD$^{-1}$ for Segments 1 \& 2.  This suggests that the superhumps in Segment 3 are somewhat less coherent than in the other two segments, in turn suggesting that the second outburst (but not the first) disrupted the nodal disk precession believed to give rise to negative superhumps \citep[e.g.][]{Harvey_NSH}.

\subsection{Eclipse Analysis}
\label{sec:ecl}
\par The luminosity $L$ of an accretion disk depends on both the temperature at each annulus in the disk ($T(r)$), which in turn depends on the local accretion rate $\dot{m}$ at each location.  For a static disk, $L$ can be given by:
\begin{equation}
L\propto\int_{R_{in}}^{R_{out}} T(r)^4r \mathrm{d}r\label{eq:1}
\end{equation}
where $R_{in}$ and $R_{out}$ are the inner and outer radii of the disk, and $T(r)$ is given by:
\begin{equation}
T(r)\propto \left(\frac{\dot m}{r^3}\left[1-\sqrt{\frac{R^*}{r}}\right]\right) ^{\frac{1}{4}}
\end{equation}
where $R^*$ is the radius of the compact object.
\par In eclipsing systems, a portion of the disk flux is obscured by the companion star during each eclipse.  In the case of a maximal eclipse, in which the companion star passes directly in front of the compact object, the total unobscured flux $\phi_\mathrm{ecl}$ from the system can be given by:
\begin{equation}
\Phi_{\rm ecl}\propto
\begin{dcases}
\Phi_0+k\int_{R_{\rm ecl}}^{R_{\rm out}} T(r)^4r \mathrm{d}r&\text{if }R_{\rm ecl}<R_{\rm out}\\
\Phi_0&\text{otherwise}
\end{dcases}
\label{eq:3}
\end{equation}
where $R_{\rm ecl}$ is the radius of the eclipsing companion star, $\Phi_0$ is the residual flux after the entire accretion disk and the compact object itself are obscured, and $k$ is a constant to convert the luminosity of the disk into the flux from the disk as seen from Earth.  $\Phi_0$ is assumed to be entirely from the companion star and constant.  As can be seen from equations \ref{eq:1} \& \ref{eq:3}, measuring the flux from an eclipsing system both in and out of eclipse allows us to break the degeneracy between the accretion rate in the system and the characteristic radii of the disk.

\par To find whether eclipses in EX Dra should cover a significant portion of the disk, we can compare the radius of the companion star to the `circularisation radius' $R_{\rm circ}$ of the disk; the smallest value of $R_{\rm out}$ corresponding to the lowest value of accretion rate that would form such a disk \citep[e.g.][]{Frank_Timescales}:
\begin{equation}
R_{\rm circ}=a(1+q)(0.5-0.277\log{q})^4
\end{equation}
where $a$ is the semi-major axis of the binary orbit and $q$ is the ratio of the donor mass to the accretor mass.  Previous studies \citep{Barwig_EXDra,Billington_EXDra} have estimated values of $q=0.795\pm0.082$, leading to a circularization radius $R_{\rm circ}\approx 0.14a$ (see also \citealp{Baptista_Spiral,Joergens_EXDraComponents}).
\par There are a number of published methods to estimate the effective radius $R_{\rm lobe}$ of a Roche lobe \citep[e.g.][]{Kopal_Roche,Paczynski_Roche_Size}, and hence the radius of a Roche-lobe filling star as is present in AWDs.  \citet{Eggleton_Roche} showed that:
\begin{equation}
R_{\rm lobe}=\frac{0.49q^{2/3}a}{0.6q^{2/3}+\ln{(1+q^{1/3})}}
\end{equation}
agrees with numerical calculations of Roche Lobe radii to within 1\% for all values of $q$.  Using this equation and the approximation of $q$ found by \citeauthor{Billington_EXDra}, we find that, in EX Dra, $R_{\rm lobe}\approx 0.36a$.  As this is a factor of $>2$ larger than $R_{\rm circ}$, we find that it would be possible for the disk to be fully eclipsed in EX Dra.
\par
\begin{figure*}
    \includegraphics[width=1\columnwidth, trim = 3mm 0mm 15mm 10mm,clip]{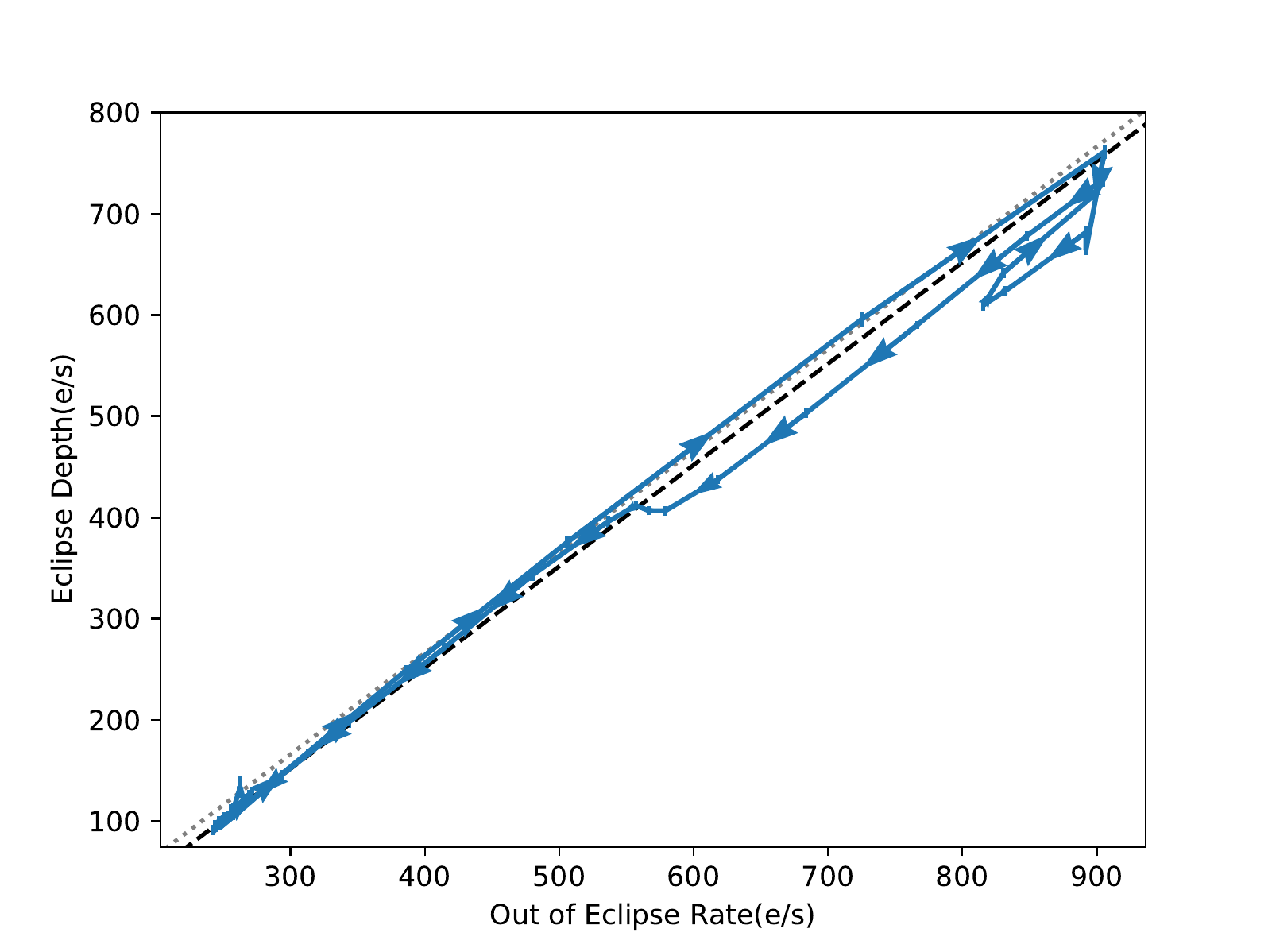}
    \includegraphics[width=1\columnwidth, trim = 3mm 0mm 15mm 10mm,clip]{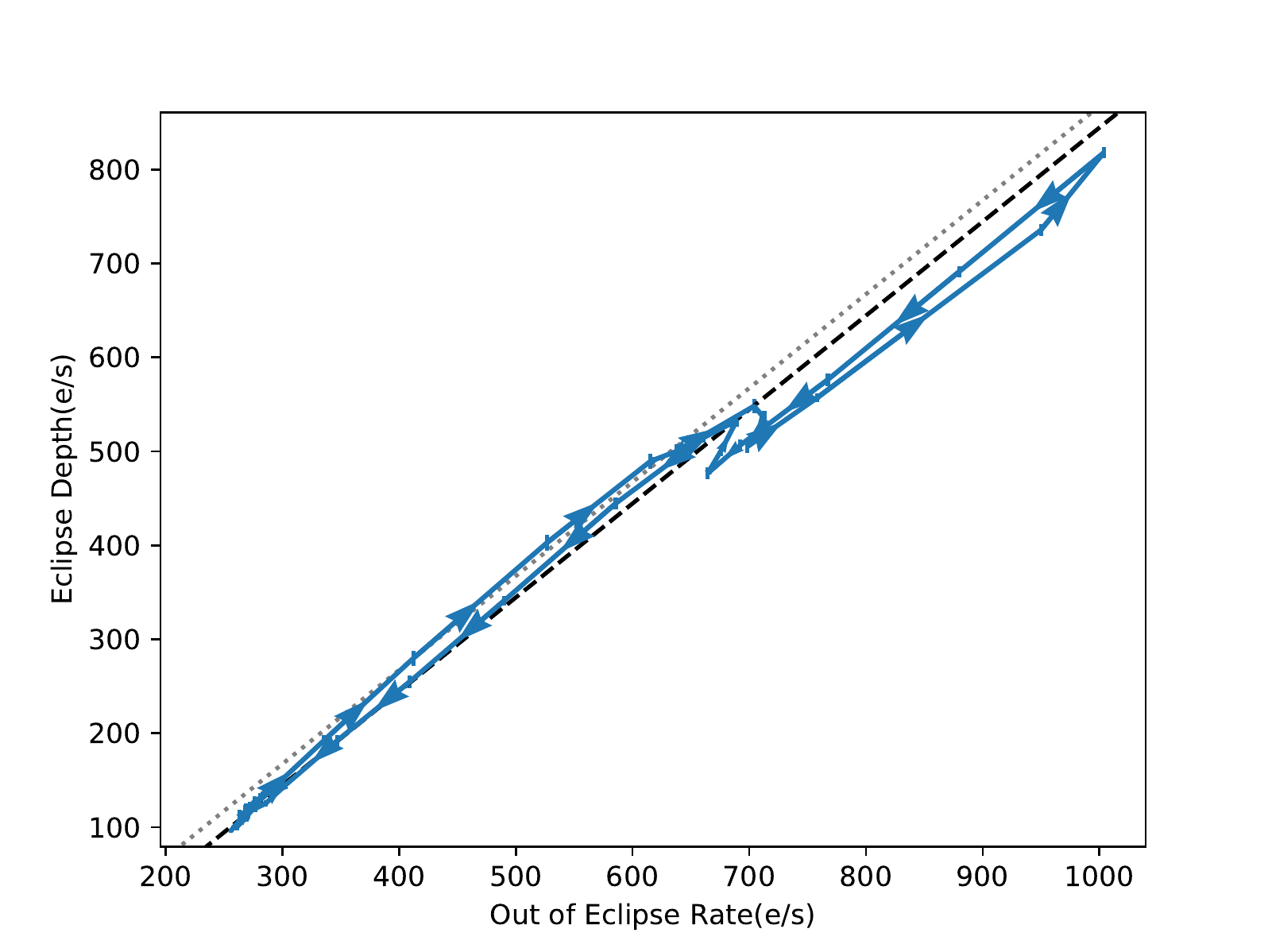}
    \captionsetup{singlelinecheck=off}
    \caption{Plots of eclipse depth against out-of-eclipse flux for the eclipses in Sector 14 (left) and Sector 15 (right).  To show the hysteretic behaviours in this parameter space, we show the line that would connect these eclipses as the source evolves in time.  In both panels, arrows indicate the direction in which the outburst progressed.  The darker dashed lines are the estimated lines of total eclipse we find by fitting a 1:1 line to the eclipses during quiescence in each sector.  The paler dotted lines indicate the estimated lines of total eclipse we find by asserting that no eclipse may have an eclipse depth that lies more than 1\,$\sigma$ above the line.  The hysteretic behaviour of the source in this parameter space can be more clearly seen when plotting fractional eclipse depth: see Figure \ref{fig:flattened}.}
   \label{fig:loops}
\end{figure*}

 In Figure \ref{fig:loops} we show plots of how the eclipse depth and the out-of-eclipse flux of EX Dra vary over the course of both of the outbursts in this study.  In both cases, we show lines with equations given by:
\begin{equation}
d_{\rm max}=\Phi-\Phi_0 
\end{equation}
where $d_{\rm max}$ is the maximum eclipse depth for a given uneclipsed flux, $\Phi$ is the estimated uneclipsed flux at that time.  To find $\Phi_0$, we fit this function to the datapoints corresponding to eclipses which occurred during quiescence.  If the radius of the disk is smaller than the radius of the companion star at this time (as found spectroscopically by \citealp{Billington_EXDra}), then datapoints on this line will then correspond to eclipses in which the entire disk is obscured \citep[e.g.][]{Scaringi_KIS}.  As the response drift of \textit{TESS} is not well described, we perform this fit to find $\Phi_0$ independently for each sector as recommended in the \textit{TESS} Instrument Handbook\footnote{\url{https://archive.stsci.edu/missions/tess/doc/TESS_Instrument_Handbook_v0.1.pdf}}.

\par Notably in each sector we find eclipses above the line of total eclipse when using our fit values of $\Phi_0$; this indicates non-physical fractional eclipse depths of greater than 100\%, in turn indicating that the accretion disk was \textit{not} fully eclipsed during quiescence.  As such, for further analysis, we redefine $\Phi_0$ in each sector as the largest value such that no rebinned\footnote{In order to reduce outlier effects on our results, we `rebin' our eclipses by a factor 2, by finding the mean depth and uneclipsed flux values of each consecutive pair of eclipses in time.} eclipse has a depth greater than $1\sigma$ above the line of total eclipse.  This new value is by definition an upper limit on $\Phi_0$, but decreasing this value further does not qualitatively change any of the results of this study.
\par
\begin{figure*}
    \includegraphics[width=1\columnwidth, trim = 3mm 0mm 15mm 10mm,clip]{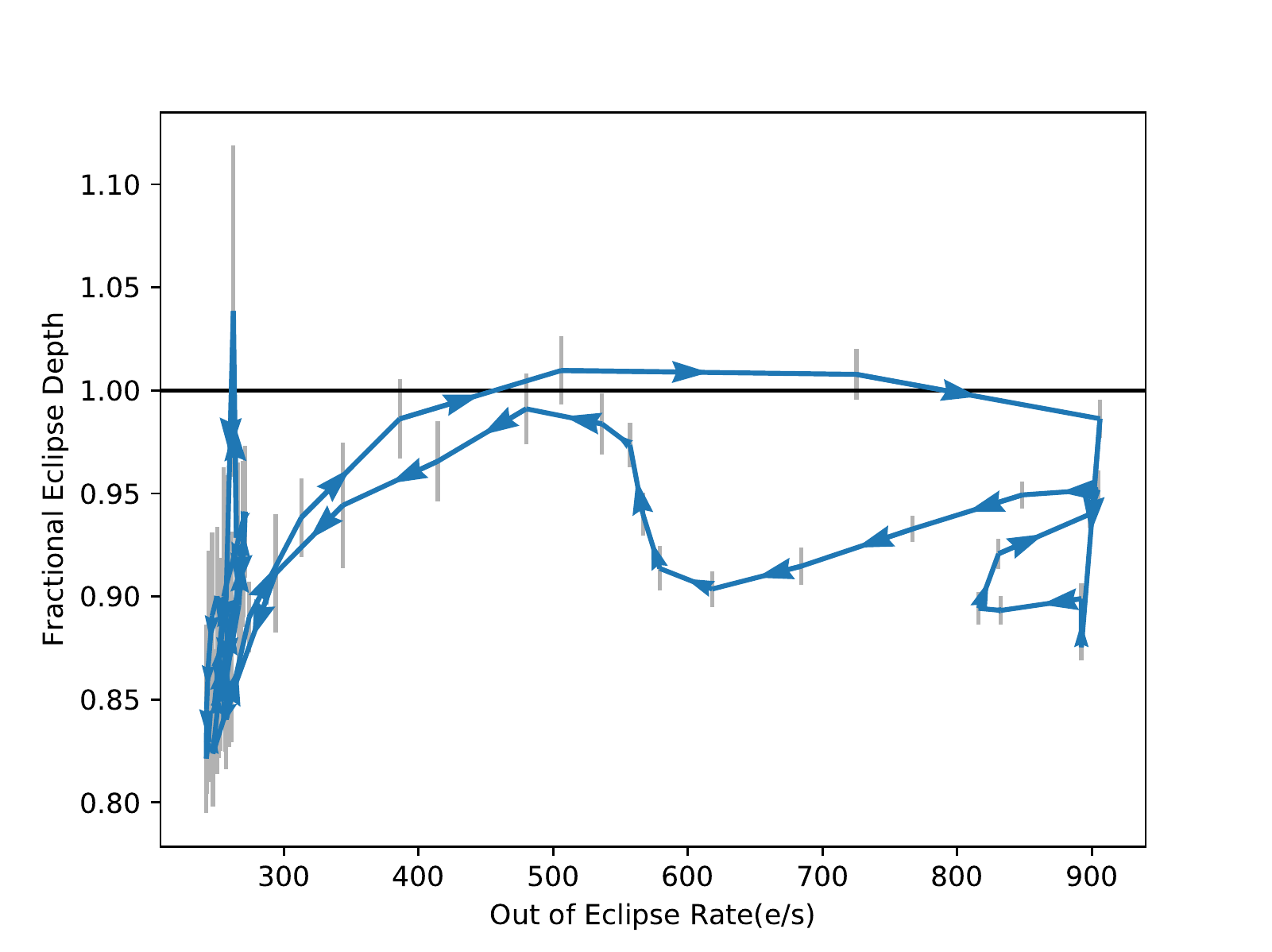}
    \includegraphics[width=1\columnwidth, trim = 3mm 0mm 15mm 10mm,clip]{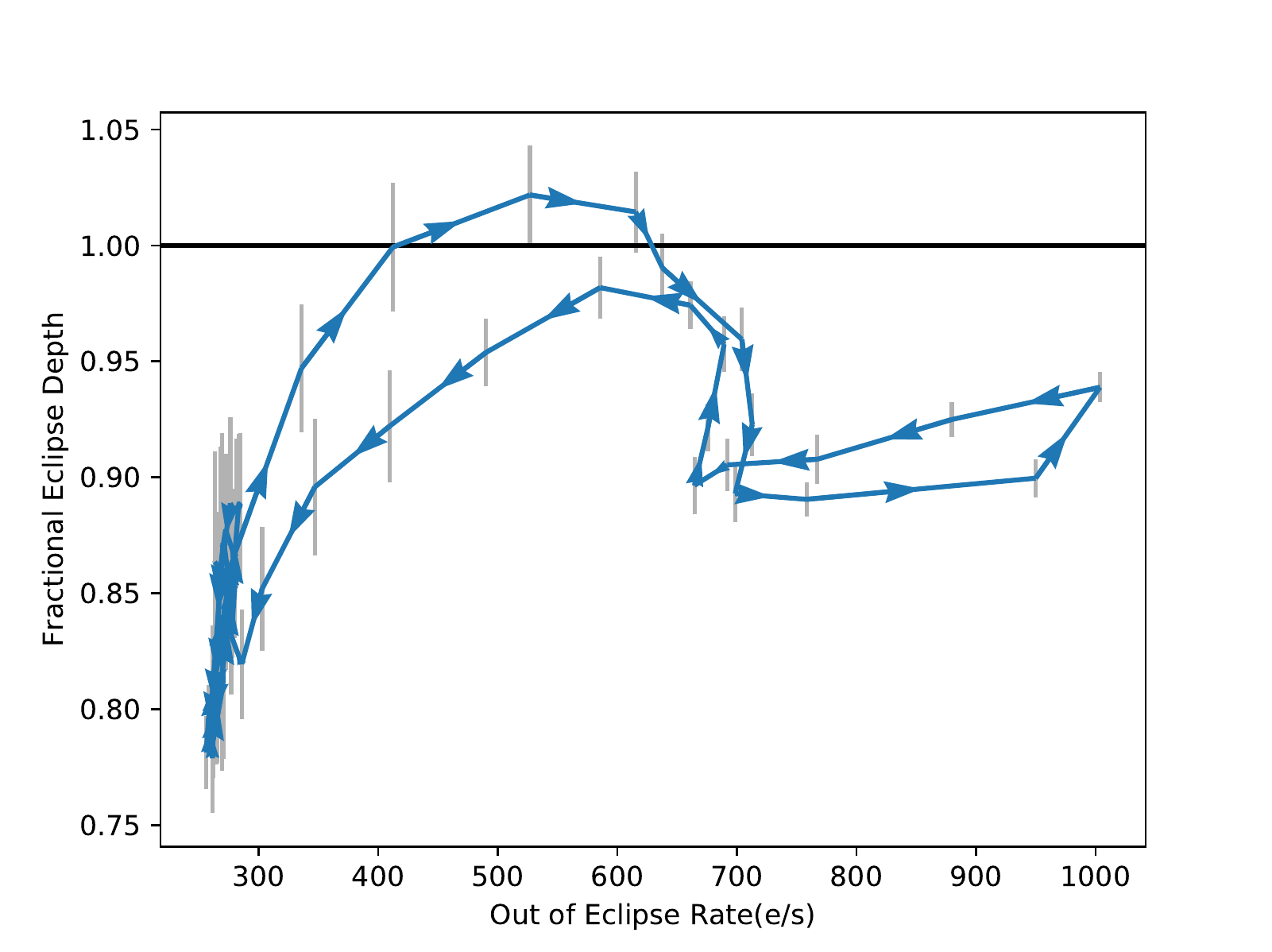}
    \captionsetup{singlelinecheck=off}
    \caption{Plots of fractional eclipse depth against out-of-eclipse flux for the eclipses in Sectors 14 (left) and 15 (right).  Arrows indicate the direction in which the hysteretic loops were executed over the course of the outburst in each Sector.  To convert eclipse depth to fractional eclipse depth we assume $\Phi_0=133.827$\,e$^-$s$^{-1}$ in Sector 14 and $\Phi_0=132.770$\,e$^-$s$^{-1}$ in Sector 15, chosen to ensure that no eclipse has an apparent fractional depth more than 1\,$\sigma$ above 1.0.}
   \label{fig:flattened}
\end{figure*}

 In both Sectors, the eclipses show significant deviation below the line of maximum eclipse during outburst, although the magnitude of this deviation is significantly smaller than those seen in eclipses of the CVs KIS J192748.53+444724.5 and Z Cha \citep{Scaringi_KIS,Court_ZCha}.  In order to better quantify the hysteresis shown in this parameter space, we can divide the depth of each eclipse by the predicted maximum eclipse depth, thus obtaining the fractional depth $\Delta$ of each eclipse:
\begin{equation}
\Delta=\frac{d}{\Phi-\Phi_0 }
\end{equation}
We show plots of fractional eclipse depth against out-of eclipse flux in Figure \ref{fig:flattened}; this figure better shows the presence and direction of hysteretic loops in this parameter space (hereafter \delphi\ space), but the exact shapes of the loops in this figure depend on the value we obtain for $\Phi_0$.  We show that, in both Sectors, eclipses during the rise of the outburst show an increase in fractional eclipse depth until they approach the line of total eclipse.  In Sector 14, eclipses remain close to the estimated line of total eclipse until near the peak of the outburst, at which point they decrease to an eclipse fraction of $\sim0.85$.  During the latter stages of the outburst, the eclipses generally have lower eclipse fractions than during the outburst rise, resulting in a generally clockwise\footnote{`Clockwise' and `anticlockwise' hysteretic loops in this study refer to the direction of a loop traced on a diagram with out-of-eclipse flux on the $x$-axis and eclipse depth on the $y$-axis.} hysteretic loop in the parameter space as defined here.  Notably, this is different to the behaviour seen in Z Cha, in which eclipses execute a generally anticlockwise hysteretic loop in the same parameter space over the course of an outburst \citep{Court_ZCha}.
\par The hysteretic behaviour in \delphi\ space in Sector 15 is broadly similar to that in Sector 14; the fractional eclipse depth tends to be smaller during the fall of the eclipse than during the rise, again leading to a generally clockwise hysteretic loop.  There is some more complex behaviour near the peak of the outburst in Sector 15 during which the eclipses trace a smaller, anticlockwise hysteretic loop.  This behaviour is likely linked to the more complex profile of the outburst in Sector 15 compared to Sector 14 (compare e.g. Figure \ref{fig:loops}).

\section{Discussion}

\par By studying the properties of eclipses in EX Dra, we have discovered the existence of hysteretic loops in eclipse-depth/out-of-eclipse space (\delphi\ space).  We have also presented evidence of negative superhumps during quiescence in this system.  Both of these results place constraints on the geometry of the EX Dra system and the physics of its accretion disk.  In this section, we discuss these constraints, and compare our findings in EX Dra to previous studies of similar objects, as well as to models of Inside-Out and Outside-In outbursts in AWDs.

\subsection{Comparison with Z Cha}

\par The different hysteretic loop structure traced by the eclipses in these outbursts highlight a number of significant differences between EX Dra and Z Cha, another AWD on which similar analysis has been performed.  \citet{Court_ZCha} found that eclipses in Z Cha underwent anticlockwise hysteretic loops in \delphi\ space during the course of both an outburst and a superoutburst.  \citeauthor{Court_ZCha} used this behaviour to attempt to understand the behaviour of the accretion disk during these events.  In order to replicate the loop seen in data, they found that the outbursts in Z Cha must begin with an increase in the physical size of the emitting region of the disk followed, after some finite time delay, by an increase in the mass transfer rate through the disk and hence its surface brightness.  The opposite direction of the loop we find here (compare Figures \ref{fig:loops} \& \ref{fig:flattened} with Figures 12 \& 13 in \citealp{Court_ZCha}) indicates that the order of events during the outburst in EX Dra is reversed; during both outbursts covered in this study, the disk first underwent an increase in surface brightness, followed by a radial increase in the size of the emitting region.
\par EX Dra and Z Cha have significantly differing orbital periods; EX Dra has an orbital period of $\sim5.04$\,hr \citep{Baptista_EXDra}, while Z Cha has an orbital period of $\sim1.79$\,hr (e.g. \citealp{McAllister_Zeclipses}).  Notably, this places the two systems on opposite sides of the so-called `period gap'; a range of orbital periods between $\sim2.2$ and $\sim2.8$\,hr which relatively few AWDs are found to possess \citep[e.g.][]{Paczynski_PeriodGap}.  A semi-empirical model of the evolution of AWDs created by \citet{Knigge_Donors} suggests that the physics of mass transfer in these objects should differ between systems above the period gap and systems below it.  In both cases, the accretion from the Roche-lobe filling donor star through L1 must be balanced by the shrinking of the orbital period of the AWD.  In systems below the period gap, \citet{Knigge_Donors} found that the system evolution is dominated by the emission of gravitational waves, leading to a slow orbital shrinkage and a corresponding low accretion rate $\dot{M}$.  Above the period gap the Roche Lobe is large enough to contain a star with a radiative core, and the main evolutionary mechanism becomes magnetic braking \citep{Rappaport_MB}.  This method of angular momentum transfer is significantly more efficient than gravitational wave emission, and hence \citet{Knigge_Donors} find that AWDs above the period gap should have higher $\dot{M}$ than systems below the gap.  As such, we would expect EX Dra to have a significantly greater $\dot{M}$ than Z Cha.
\par Another notable phenomenological difference between Z Cha and EX Dra is the behaviour of the eclipses during quiescence and at the start and end of each outburst.  In Z Cha, \citet{Court_ZCha} found that eclipses during quiescence were consistent with having fractional depths of $100\%$, and that this remained the case during the initial stages of each outburst.  In EX Dra, we have shown that the fractional eclipse depth actually \textit{increases} during the initial stage of the outburst, meaning that the eclipse fraction during quiescence must have been $<100\%$.  The differences between Z Cha and EX Dra, in terms of how their eclipses properties vary over the course of an outburst, suggest that the evolutions of the outbursts observed in each source evolve in different ways.

\subsection{Distinguishing Inside-Out and Outside-In Outbursts}
\label{sec:IO}
\par The different ways in which an accretion disk evolves in both an Inside-Out and an Outside-In outburst is reflected in the behaviour of a source's hysteretic behaviour in \delphi\ space.  In an Outside-In outburst, the outburst onset is preceded by a build-up of material at the outer edge of the disk, so it should be expected that the outer radius of the disk increases before its temperature.  This is consistent with the results \citet{Court_ZCha} obtained by studying the eclipses in Z Cha, suggesting that the outbursts in that source are indeed of the Outside-In type.  We do not however see this behaviour in the eclipses of EX Dra, suggesting that the outbursts from this source are \textit{not} Outside-In in nature.

\begin{figure*}
    \includegraphics[width=2.0\columnwidth, trim = 0mm 0mm 0mm 0mm]{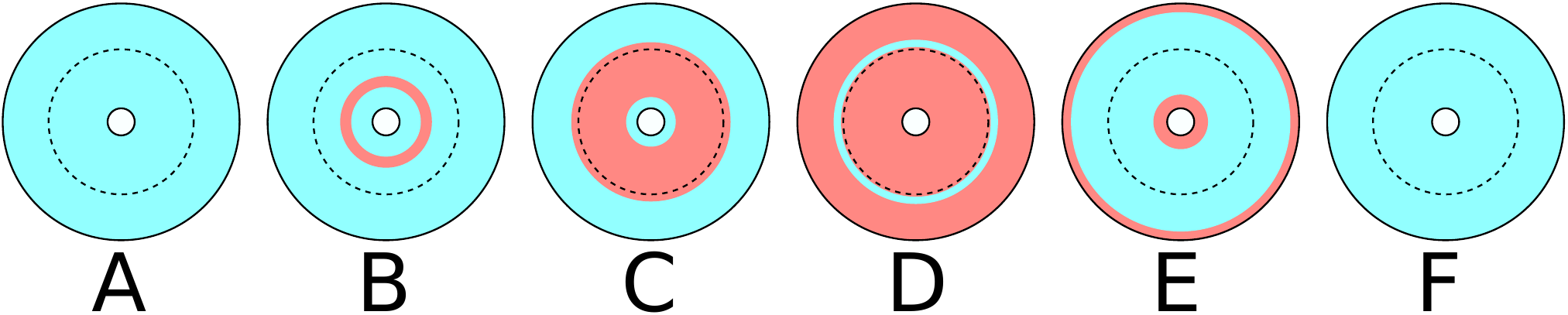}
    \includegraphics[width=2.0\columnwidth, trim = 18mm 0mm 14mm 10mm,clip]{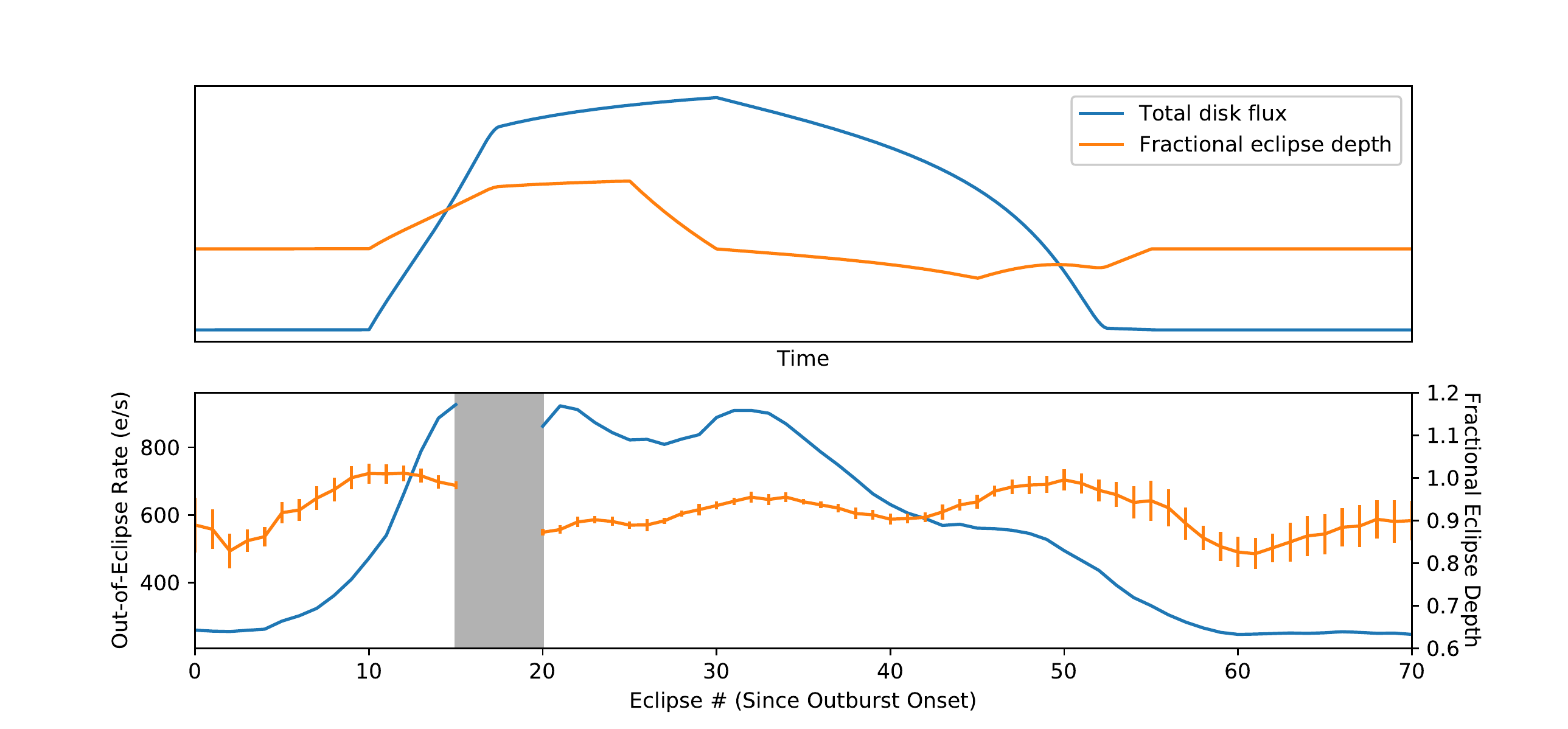}
    \captionsetup{singlelinecheck=off}
    \caption{\textbf{Upper:} A diagram representing the evolution of an accretion disk during an Inside-Out outburst as described by \citet{Mineshige_OIIO}.  In each image, portions of the disk in the `cold' state are shaded in blue, portions of the disk in the `hot' state are shaded in red and the radius $r_{e}$ corresponding to the radius of the eclipsing red dwarf is shown as a dashed line.  In stage \textbf{B} a heating wave is triggered in the inner portion of the disk, within $r_e$.  This raises the fractional eclipse depth until stage \textbf{C} when the heating wave spreads beyond $r_e$ and the fractional eclipse depth begins to decrease again.  In stage \textbf{D} a cooling wave is triggered at some radius outside that at which the heating wave was triggered, proagating throughout the disk (\textbf{E}) and eventually returning the disk to a quiescent state \textbf{F}.  The accretion disks shown are not to scale.  \textbf{Middle:} a plot, from a simple model based on this evolution and the disk temperature profile solutions of \citet{Shakura_Disk}, showing how the total disk luminosity and fractional eclipse depth would vary throughout such an outburst.  \textbf{Lower:} The out-of-eclipse rate and fractional eclipse depths which we infer for the outburst in Sector 14, to be compared with the predicted behaviour of these parameters in a simple inside-out outburst.  The grey shaded region corresponds to the data gap in this sector.}
   \label{fig:ecldiag}
\end{figure*}

\par If we assume that the accretion disk in EX Dra is indeed larger than the companion star during quiescence, then an Inside-Out outburst can explain the increase in eclipse fraction during the outburst rise; Inside-Out outbursts have previously been identified in EX Dra during the 1990s \citep{Baptista_EXDra}.  At the onset of such an outburst, only the inner region of the disk experiences an increase in surface brightness, while the luminosity of the outer disk remains unchanged (shown in stages \textbf{A}$\rightarrow$\textbf{B} in Figure \ref{fig:ecldiag}).  Assuming that the `extra luminosity' $L_e$ due to some portion of the disk going to the hot state is entirely concentrated in the region of the disk which is eclipsed, we find:
\begin{equation}
\label{eq:increaser}
\Delta=\frac{d_q+L_e}{L_q+L_e}
\end{equation}
where $d_q$ and $L_q$ are the eclipse depth and out-of-eclipse disk flux when the disk is fully in the cold state during quiescence.  Assuming $L_q>d_q$, i.e. some part of the disk remains uneclipsed during quiescence, we see that $\Delta$ increases as the surface brightness of the inner disk increases.
\par As the outburst progresses, the outer radius of the hot region of the disk increases and Equation \ref{eq:increaser} no longer holds (Stage \textbf{C} in Figure \ref{fig:ecldiag}).  As more of the uneclipsed portion of the disk also goes into the hot state, the eclipse fraction decreases again, and the eclipses in a \delphi\ plot diverge from line of maximum eclipse.  The outwards heating wave will reach the outer edge of the disk or dissipate, marking the peak of the outburst.  It can be shown that a cooling wave following behind the outwards heating wave will then begin to propagate inwards and outwards from a radius midway through the disk (Stages \textbf{D}$\rightarrow$\textbf{F} in Figure \ref{fig:ecldiag}), extinguishing the outburst and returning the disk to the cool state \citep{Mineshige_OIIO}.  During this return to quiescence, the emission from the disk is less centrally peaked than during the outburst rise.  This leads to eclipse fractions being generally lower during the outburst decay than during the rise, in turn giving rise to a clockwise hysteretic loop in \delphi\ space much as we see in Figure \ref{fig:flattened}.
\par Due to the similarities between the data and what would be expected from the scenarios described above, we conclude that the outbursts in both Sectors 14 and 15 were of the Inside-Out type.  The models of \citet{Mineshige_OIIO} predict that Inside-Out outbursts should only occur in AWDs with a low accretion rate, in turn suggesting that the accretion rate in EX Dra is relatively low.  This result is at odds with the expectation that systems above the period gap, such as EX Dra, have significantly \textit{higher} accretion rates \citep{Knigge_Donors} than those below the period gap (such as Z Cha, a system which we interpret as undergoing Outside-In outbursts).  Our result serves as further evidence that global $\dot{M}$ is not the sole criteria separating inside-out and Outside-In outbursts in AWDs (see also \citealp{BuatMenard_NovaModelFix}).
\par The rise in $\Delta$ during the onset of the outburst is dependent on the disk not being fully eclipsed during quiescence.  This contradicts the results of \citet{Billington_EXDra} who found that the emission during eclipses in quiescence is spectroscopically consistent with being entirely from the companion red dwarf.  This disparity is likely due to the fact that the data used by \citeauthor{Billington_EXDra} was taken solely in the H$\alpha$ band; in this band, the relatively red companion star contributes a much higher fraction of the system's flux, making it easier to reduce the disk flux to an undetectable fraction during eclipses.  In any case, we find that the hysteretic loop in \delphi\ space is \textit{not} dependent on the quiescent radius of the accretion disk; an outburst as described above should always show a clockwise loop in this parameter space regardless of whether the disk is smaller or larger than the eclipsing companion at the start of the outburst.  This is a strong contrast with the anticlockwise hysteretic loop observed during Outside-In outbursts (e.g. \citealp{Court_ZCha}).  As such, we propose that the direction of a hysteretic loop in \delphi\ space can be used as a purely photometric test to differentiate between inside-out and Outside-In outbursts in eclipsing AWDs.

\subsection{Outburst onset in Tilted Accretion Disks}

\par In systems with high accretion rates, matter gathers at the outer edge of the disk faster that it can be transferred inwards, and the critical surface density is triggered at the outer edge of the disk, triggering Outside-In outbursts.  As such, Outside-In outbursts are believed to be the more common type of outburst in high accretion systems, with Inside-Out outbursts more common in low accretion rate systems \citep[e.g.][]{Mineshige_OIIO}.  The accretion rate in EX Dra is not known, but the well-constrained $\sim5.03$\,hr period of the system places it firmly above the `period gap', a range of periods between 2--3\,hr in which relatively few CVs are found.  Above the period gap the inspiral rate of the binary, and hence the mass transfer rate $\dot{M}$, is driven by magnetic braking.  However, below the period gap, it can be shown that a Roche-lobe filling near-main sequence star should be fully convective in nature, and hence not support a significant magnetic field.  In these systems, the inspiral rate is instead driven by the much slower process of the emitting gravitational radiation.  As such it is expected that, in general, AWDs above the period gap should have higher $\dot{M}$ than systems below it \citep{Rappaport_MB0,Rappaport_MB,Spruit_Gap}.
\par As EX Dra is above the period gap, the system is likely to have a relatively high accretion rate.  This is seemingly at odds with the fact that this sytem consistently undergoes Inside-Out outbursts (e.g. \citealp{Baptista_EXDra}, our discussion in Section \ref{sec:IO}).  However, the models of \citet{Mineshige_OIIO} are calculated for a smoothly accreting circular disk in the orbital plane of the system, and a number of other authors have proposed ways to trigger Inside-Out outbursts in more complex high-$\dot{M}$ systems.  For example, magnetohydronamic simulations by \citet{Armitage_Overflow} have shown that, in systems with very high $\dot{M}$, a portion of the accretion flow can ricochet off the outer edge of the disk and proceed in an arc to accrete directly onto the inner disk.  However, it is unclear whether the $\dot{M}$ in EX Dra is large enough for this process to occur.
\par The presence of superhumps in EX Dra however does suggest that this system is also more geometrically complex than the systems modelled by \citet{Mineshige_OIIO}.  Negative superhumps, such as we see in this system, are believed to be caused by a beat between the orbital frequency of a system and a slow vertical precession of a tilted accretion disk \citep[e.g.][]{Bonnet-Bidaud_NegativeSuperhumps,Patterson_NegativeSuperhump}.  As such the presence of negative superhumps, such as we find in EX Dra, is evidence for such a tilted accretion disk.
\par Recent simulations by \citet{Kimura_Tilt} have shown that, in AWD systems containing a highly tilted accretion disk, some of the matter donated from the companion star can bypass the outer accretion disk and fall directly onto the inner regions of the disk.  This causes the surface density in the inner regions of the accretion disk to increase disproportionately quickly compared to the outer regions.  This in turn should lead to inside-out outbursts being the dominant form of outbursts in such system.  As such, the presence of negative superhumps in EX Dra, as well as our identification of inside-out outbursts in this source, provides direct evidence for the accretion scenario proposed by \citeauthor{Kimura_Tilt}.  This in turn can explain how inside-out outbursts can dominate in a system which likely has a high accretion rate.

\section{Conclusions}

\par We have performed a study of the eclipses of the AWD EX Dra during the first two outbursts which were observed by \textit{TESS}.  We have analysed how these eclipses evolve in eclipse-depth/out-of-eclipse-flux space (\delphi\ space) over the course of each outburst, finding significant hysteretic behaviour in both cases.  This makes EX Dra the third AWD in which such hysteresis has been quantified, after KIS J192748.53+444724.5 \citep{Scaringi_KIS} and Z Cha \citep{Court_ZCha}, strengthening the argument that this behaviour is likely common to all eclipsing dwarf nova systems.  Eclipse depth variability over the course of an outburst has also been reported in a number of other eclipsing CVs (e.g. V447 Lyr, \citealp{Ramsay_PhaseJump}, V729 Sgr, \citealp{Ramsay_729}, and CRTS CRTS J035905.9+175034, \citealp{Littlefield_PhaseJump}), suggesting that there are many more systems to which these methods can be applied.
\par We find that during the initial stages of each outburst, the fractional eclipse depth rises, in turn implying that the accretion disk cannot have been fully eclipsed during quiescence.  This is a direct contradiction to the results of a spectroscopic study performed by \citet{Billington_EXDra}, which is likely due to differences in the instrumentation and energy bands used in our respective studies.
\par We have found that the hysteretic loops in \delphi\ space in EX Dra are executed in the opposite direction to those seen in Z Cha.  We have shown that this difference would be expected if the outbursts in EX Dra are of the `Inside Out' type (outbursts that begin in the inner regions of the accretion disk) as opposed to the `Outside In' outbursts seen in Z Cha, despite EX Dra being expected to have a higher accretion rate than Z Cha.  As such, we have found that the direction of the hysteretic loops in \delphi\ space can be used as a reliable and purely photometric method to phenomenologically distinguish between the two types of outburst.
\par As analysis of hysteretic \delphi\ loops can be performed on any dwarf nova system which shows deep eclipses, this new method gives us a way to categorise historical dwarf nova outbursts in archival data, as well as any future outbursts which occur in these systems.  This will significantly increase the sample size of outbursts which are reliably known to belong to each type (`Inside-Out' or `Outside In'), allowing for population studies into how AWD phenomenology differs between these two types of outburst.  As AWDs are oriented randomly in space, we can expect $\sim11\%$ to be at angles of $\gtrsim85^\circ$, and hence likely eclipsors, meaning that the total sample of AWDs on which our method can be applied should be high.  Applying our method to identify inside-out outbursts to additional AWDs which show negative superhumps will also be able to provide additional evidence for the model of accretion proposed by \citet{Kimura_Tilt}, in which an accretion stream feeds directly into the inner portions of a highly tilted disk.

\section*{Acknowledgements}

\par In this study, we make use of the Numpy, Scipy \citep{Numpy} and Astropy \citep{Astropy} libraries for Python.  Figures in this paper were produced using MatplotLib \citep{Hunter_MatPlotLib}.  This paper includes data collected with the \textit{TESS mission}, obtained from the MAST data archive at the Space Telescope Science Institute (STScI). Funding for the \textit{TESS} mission is provided by the NASA Explorer Program. STScI is operated by the Association of Universities for Research in Astronomy, Inc., under NASA contract NAS 5--26555.  SS and JMCC acknowledge support for this work from the TESS Guest Investigator program under NASA grant 80NSSC19K1735.  DA acknowledges support from the Royal Society.  ND is supported by a Vidi grant from the Netherlands Organization for Scientific Research (NWO).

%The Acknowledgements section is not numbered. Here you can thank helpful
%colleagues, acknowledge funding agencies, telescopes and facilities used etc.

%%%%%%%%%%%%%%%%%%%%%%%%%%%%%%%%%%%%%%%%%%%%%%%%%%



\bibliographystyle{mnras}
\bibliography{refs}

\newcommand{\noop}[1]{}
\begin{thebibliography}{}
\makeatletter
\relax
\def\mn@urlcharsother{\let\do\@makeother \do\$\do\&\do\#\do\^\do\_\do\%\do\~}
\def\mn@doi{\begingroup\mn@urlcharsother \@ifnextchar [ {\mn@doi@}
  {\mn@doi@[]}}
\def\mn@doi@[#1]#2{\def\@tempa{#1}\ifx\@tempa\@empty \href
  {http://dx.doi.org/#2} {doi:#2}\else \href {http://dx.doi.org/#2} {#1}\fi
  \endgroup}
\def\mn@eprint#1#2{\mn@eprint@#1:#2::\@nil}
\def\mn@eprint@arXiv#1{\href {http://arxiv.org/abs/#1} {{\tt arXiv:#1}}}
\def\mn@eprint@dblp#1{\href {http://dblp.uni-trier.de/rec/bibtex/#1.xml}
  {dblp:#1}}
\def\mn@eprint@#1:#2:#3:#4\@nil{\def\@tempa {#1}\def\@tempb {#2}\def\@tempc
  {#3}\ifx \@tempc \@empty \let \@tempc \@tempb \let \@tempb \@tempa \fi \ifx
  \@tempb \@empty \def\@tempb {arXiv}\fi \@ifundefined
  {mn@eprint@\@tempb}{\@tempb:\@tempc}{\expandafter \expandafter \csname
  mn@eprint@\@tempb\endcsname \expandafter{\@tempc}}}

\bibitem[\protect\citeauthoryear{{Armitage} \& {Livio}}{{Armitage} \&
  {Livio}}{1996}]{Armitage_Overflow}
{Armitage} P.~J.,  {Livio} M.,  1996, \mn@doi [\apj] {10.1086/177928}, \href
  {https://ui.adsabs.harvard.edu/abs/1996ApJ...470.1024A} {470, 1024}

\bibitem[\protect\citeauthoryear{{Astropy Collaboration} et~al.,}{{Astropy
  Collaboration} et~al.}{2013}]{Astropy}
{Astropy Collaboration} et~al., 2013, \mn@doi [\aap]
  {10.1051/0004-6361/201322068}, \href
  {http://adsabs.harvard.edu/abs/2013A%26A...558A..33A} {558, A33}

\bibitem[\protect\citeauthoryear{{Baptista} \& {Catal{\'a}n}}{{Baptista} \&
  {Catal{\'a}n}}{2000}]{Baptista_Spiral}
{Baptista} R.,  {Catal{\'a}n} M.~S.,  2000, \mn@doi [\apjl] {10.1086/312834},
  \href {https://ui.adsabs.harvard.edu/abs/2000ApJ...539L..55B} {539, L55}

\bibitem[\protect\citeauthoryear{{Baptista}, {Catal{\'a}n}  \&
  {Costa}}{{Baptista} et~al.}{2000}]{Baptista_EXDra}
{Baptista} R.,  {Catal{\'a}n} M.~S.,   {Costa} L.,  2000, \mn@doi [\mnras]
  {10.1046/j.1365-8711.2000.03557.x}, \href
  {https://ui.adsabs.harvard.edu/abs/2000MNRAS.316..529B} {316, 529}

\bibitem[\protect\citeauthoryear{{Barwig}, {Ritter}  \& {Barnbantner}}{{Barwig}
  et~al.}{1994}]{Barwig_EXDra}
{Barwig} H.,  {Ritter} H.,   {Barnbantner} O.,  1994, \aap, \href
  {https://ui.adsabs.harvard.edu/abs/1994A&A...288..204B} {288, 204}

\bibitem[\protect\citeauthoryear{{Billington}, {Marsh}  \&
  {Dhillon}}{{Billington} et~al.}{1996}]{Billington_EXDra}
{Billington} I.,  {Marsh} T.~R.,   {Dhillon} V.~S.,  1996, \mn@doi [\mnras]
  {10.1093/mnras/278.3.673}, \href
  {https://ui.adsabs.harvard.edu/abs/1996MNRAS.278..673B} {278, 673}

\bibitem[\protect\citeauthoryear{{Bonnet-Bidaud}, {Motch}  \&
  {Mouchet}}{{Bonnet-Bidaud} et~al.}{1985}]{Bonnet-Bidaud_NegativeSuperhumps}
{Bonnet-Bidaud} J.~M.,  {Motch} C.,   {Mouchet} M.,  1985, \aap, \href
  {https://ui.adsabs.harvard.edu/abs/1985A&A...143..313B} {143, 313}

\bibitem[\protect\citeauthoryear{{Buat-M{\'e}nard}, {Hameury}  \&
  {Lasota}}{{Buat-M{\'e}nard} et~al.}{2001}]{BuatMenard_NovaModelFix}
{Buat-M{\'e}nard} V.,  {Hameury} J.~M.,   {Lasota} J.~P.,  2001, \mn@doi [\aap]
  {10.1051/0004-6361:20000107}, \href
  {https://ui.adsabs.harvard.edu/abs/2001A&A...366..612B} {366, 612}

\bibitem[\protect\citeauthoryear{{Court} et~al.,}{{Court}
  et~al.}{2019}]{Court_ZCha}
{Court} J.~M.~C.,  et~al., 2019, \mn@doi [\mnras] {10.1093/mnras/stz2015},
  \href {https://ui.adsabs.harvard.edu/abs/2019MNRAS.488.4149C} {488, 4149}

\bibitem[\protect\citeauthoryear{{Eggleton}}{{Eggleton}}{1983}]{Eggleton_Roche}
{Eggleton} P.~P.,  1983, \mn@doi [\apj] {10.1086/160960}, \href
  {https://ui.adsabs.harvard.edu/abs/1983ApJ...268..368E} {268, 368}

\bibitem[\protect\citeauthoryear{{Fiedler}, {Barwig}  \& {Mantel}}{{Fiedler}
  et~al.}{1997}]{Fiedler_EXDra}
{Fiedler} H.,  {Barwig} H.,   {Mantel} K.~H.,  1997, \aap, \href
  {https://ui.adsabs.harvard.edu/abs/1997A&A...327..173F} {327, 173}

\bibitem[\protect\citeauthoryear{{Frank}, {King}  \& {Raine}}{{Frank}
  et~al.}{2002}]{Frank_Timescales}
{Frank} J.,  {King} A.,   {Raine} D.~J.,  2002, {Accretion Power in
  Astrophysics: Third Edition}

\bibitem[\protect\citeauthoryear{{Golysheva}, {Shugarov}, {Katysheva}  \&
  {Khruzina}}{{Golysheva} et~al.}{2015}]{Golysheva_LCAnalysis}
{Golysheva} P.,  {Shugarov} S.,  {Katysheva} N.,   {Khruzina} T.,  2015, in
  {Rucinski} S.~M.,  {Torres} G.,   {Zejda} M.,  eds,  Astronomical Society of
  the Pacific Conference Series Vol. 496, Living Together: Planets, Host Stars
  and Binaries. pp 231--235

\bibitem[\protect\citeauthoryear{{Harvey}, {Skillman}, {Patterson}  \&
  {Ringwald}}{{Harvey} et~al.}{1995}]{Harvey_NSH}
{Harvey} D.,  {Skillman} D.~R.,  {Patterson} J.,   {Ringwald} F.~A.,  1995,
  \mn@doi [\pasp] {10.1086/133591}, \href
  {https://ui.adsabs.harvard.edu/abs/1995PASP..107..551H} {107, 551}

\bibitem[\protect\citeauthoryear{{Horne}}{{Horne}}{1984}]{Horne_Superhump}
{Horne} K.,  1984, \mn@doi [\nat] {10.1038/312348a0}, \href
  {http://adsabs.harvard.edu/abs/1984Natur.312..348H} {312, 348}

\bibitem[\protect\citeauthoryear{{Hunter}}{{Hunter}}{2007}]{Hunter_MatPlotLib}
{Hunter} J.~D.,  2007, Computing In Science \& Engineering, 9, 90

\bibitem[\protect\citeauthoryear{{Ichimaru}}{{Ichimaru}}{1977}]{Ichimaru_DiskStates}
{Ichimaru} S.,  1977, \mn@doi [\apj] {10.1086/155314}, \href
  {https://ui.adsabs.harvard.edu/abs/1977ApJ...214..840I} {214, 840}

\bibitem[\protect\citeauthoryear{{Ioannou}, {Naylor}, {Welsh}, {Catal{\'a}n},
  {Worraker}  \& {James}}{{Ioannou} et~al.}{1999}]{Ioannou_OutsideIn}
{Ioannou} Z.,  {Naylor} T.,  {Welsh} W.~F.,  {Catal{\'a}n} M.~S.,  {Worraker}
  W.~J.,   {James} N.~D.,  1999, \mn@doi [\mnras]
  {10.1046/j.1365-8711.1999.03001.x}, \href
  {https://ui.adsabs.harvard.edu/abs/1999MNRAS.310..398I} {310, 398}

\bibitem[\protect\citeauthoryear{{Irwin}, {Campbell}, {Morbey}, {Walker}  \&
  {Yang}}{{Irwin} et~al.}{1989}]{Irwin_LombScargle}
{Irwin} A.~W.,  {Campbell} B.,  {Morbey} C.~L.,  {Walker} G.~A.~H.,   {Yang}
  S.,  1989, \mn@doi [\pasp] {10.1086/132415}, \href
  {http://adsabs.harvard.edu/abs/1989PASP..101..147I} {101, 147}

\bibitem[\protect\citeauthoryear{{Joergens}, {Mantel}, {Barwig},
  {B{\"a}rnbantner}  \& {Fiedler}}{{Joergens}
  et~al.}{2000}]{Joergens_EXDraComponents}
{Joergens} V.,  {Mantel} K.-H.,  {Barwig} H.,  {B{\"a}rnbantner} O.,
  {Fiedler} H.,  2000, \aap, \href
  {https://ui.adsabs.harvard.edu/abs/2000A&A...354..579J} {354, 579}

\bibitem[\protect\citeauthoryear{{Jones}, {Oliphant}, {Peterson}
  et~al.}{{Jones} et~al.}{2001}]{Numpy}
{Jones} E.,  {Oliphant} T.,  {Peterson} P.,   et~al., 2001, {{SciPy}: Open
  source scientific tools for {Python}}, \url {http://www.scipy.org/}

\bibitem[\protect\citeauthoryear{{Kato} \& {Osaki}}{{Kato} \&
  {Osaki}}{2013}]{Kato_3DN}
{Kato} T.,  {Osaki} Y.,  2013, \mn@doi [\pasj] {10.1093/pasj/65.5.97}, \href
  {https://ui.adsabs.harvard.edu/abs/2013PASJ...65...97K} {65, 97}

\bibitem[\protect\citeauthoryear{{Khruzina}, {Voloshina}, {Qian}, {Wolf}  \&
  {Metlov}}{{Khruzina} et~al.}{2019}]{Khruzina_EXDra}
{Khruzina} T.~S.,  {Voloshina} I.~B.,  {Qian} S.,  {Wolf} M.,   {Metlov} V.~G.,
   2019, \mn@doi [Astronomy Reports] {10.1134/S1063772919070035}, \href
  {https://ui.adsabs.harvard.edu/abs/2019ARep...63..571K} {63, 571}

\bibitem[\protect\citeauthoryear{{Kimura}, {Osaki}, {Kato}  \&
  {Mineshige}}{{Kimura} et~al.}{2019}]{Kimura_Tilt}
{Kimura} M.,  {Osaki} Y.,  {Kato} T.,   {Mineshige} S.,  2019, arXiv e-prints,
  \href {https://ui.adsabs.harvard.edu/abs/2019arXiv191207217K} {p.
  arXiv:1912.07217}

\bibitem[\protect\citeauthoryear{{Knigge}}{{Knigge}}{2006}]{Knigge_Relation}
{Knigge} C.,  2006, \mn@doi [\mnras] {10.1111/j.1365-2966.2006.11096.x}, \href
  {http://adsabs.harvard.edu/abs/2006MNRAS.373..484K} {373, 484}

\bibitem[\protect\citeauthoryear{{Knigge}, {Baraffe}  \& {Patterson}}{{Knigge}
  et~al.}{2011}]{Knigge_Donors}
{Knigge} C.,  {Baraffe} I.,   {Patterson} J.,  2011, \mn@doi [\apjs]
  {10.1088/0067-0049/194/2/28}, \href
  {https://ui.adsabs.harvard.edu/abs/2011ApJS..194...28K} {194, 28}

\bibitem[\protect\citeauthoryear{{Kopal}}{{Kopal}}{1959}]{Kopal_Roche}
{Kopal} Z.,  1959, {Close binary systems}

\bibitem[\protect\citeauthoryear{{Littlefield}, {Garnavich}, {Kennedy},
  {Szkody}  \& {Dai}}{{Littlefield} et~al.}{2018}]{Littlefield_PhaseJump}
{Littlefield} C.,  {Garnavich} P.,  {Kennedy} M.,  {Szkody} P.,   {Dai} Z.,
  2018, \mn@doi [\aj] {10.3847/1538-3881/aabcd1}, \href
  {http://adsabs.harvard.edu/abs/2018AJ....155..232L} {155, 232}

\bibitem[\protect\citeauthoryear{{Lomb}}{{Lomb}}{1976}]{Lomb_LombScargle}
{Lomb} N.~R.,  1976, \mn@doi [\apss] {10.1007/BF00648343}, \href
  {http://adsabs.harvard.edu/abs/1976Ap%26SS..39..447L} {39, 447}

\bibitem[\protect\citeauthoryear{{McAllister} et~al.,}{{McAllister}
  et~al.}{2019}]{McAllister_Zeclipses}
{McAllister} M.,  et~al., 2019, \mn@doi [\mnras] {10.1093/mnras/stz976}, \href
  {http://adsabs.harvard.edu/abs/2019MNRAS.tmp.1087M} {}

\bibitem[\protect\citeauthoryear{{Meyer} \& {Meyer-Hofmeister}}{{Meyer} \&
  {Meyer-Hofmeister}}{1984}]{Meyer_Nova}
{Meyer} F.,  {Meyer-Hofmeister} E.,  1984, \aap, \href
  {http://adsabs.harvard.edu/abs/1984A%26A...132..143M} {132, 143}

\bibitem[\protect\citeauthoryear{{Mineshige} \& {Osaki}}{{Mineshige} \&
  {Osaki}}{1983}]{Mineshige_S}
{Mineshige} S.,  {Osaki} Y.,  1983, \pasj, \href
  {https://ui.adsabs.harvard.edu/abs/1983PASJ...35..377M} {35, 377}

\bibitem[\protect\citeauthoryear{{Mineshige} \& {Osaki}}{{Mineshige} \&
  {Osaki}}{1985}]{Mineshige_OIIO}
{Mineshige} S.,  {Osaki} Y.,  1985, \pasj, \href
  {https://ui.adsabs.harvard.edu/abs/1985PASJ...37....1M} {37, 1}

\bibitem[\protect\citeauthoryear{{Paczy{\'n}ski}}{{Paczy{\'n}ski}}{1971}]{Paczynski_Roche_Size}
{Paczy{\'n}ski} B.,  1971, \mn@doi [\araa]
  {10.1146/annurev.aa.09.090171.001151}, \href
  {https://ui.adsabs.harvard.edu/abs/1971ARA&A...9..183P} {9, 183}

\bibitem[\protect\citeauthoryear{{Paczynski} \& {Sienkiewicz}}{{Paczynski} \&
  {Sienkiewicz}}{1983}]{Paczynski_PeriodGap}
{Paczynski} B.,  {Sienkiewicz} R.,  1983, \mn@doi [\apj] {10.1086/161004},
  \href {https://ui.adsabs.harvard.edu/abs/1983ApJ...268..825P} {268, 825}

\bibitem[\protect\citeauthoryear{{Patterson}, {Thomas}, {Skillman}  \&
  {Diaz}}{{Patterson} et~al.}{1993}]{Patterson_NegativeSuperhump}
{Patterson} J.,  {Thomas} G.,  {Skillman} D.~R.,   {Diaz} M.,  1993, \mn@doi
  [\apjs] {10.1086/191777}, \href
  {https://ui.adsabs.harvard.edu/abs/1993ApJS...86..235P} {86, 235}

\bibitem[\protect\citeauthoryear{{Pilar{\v{c}}{\'\i}k}, {Wolf}, {Dubovsk{\'y}},
  {Hornoch}  \& {Kotkov{\'a}}}{{Pilar{\v{c}}{\'\i}k}
  et~al.}{2012}]{Pilarcik_EXDra}
{Pilar{\v{c}}{\'\i}k} L.,  {Wolf} M.,  {Dubovsk{\'y}} P.~A.,  {Hornoch} K.,
  {Kotkov{\'a}} L.,  2012, \mn@doi [\aap] {10.1051/0004-6361/201117972}, \href
  {https://ui.adsabs.harvard.edu/abs/2012A&A...539A.153P} {539, A153}

\bibitem[\protect\citeauthoryear{{Ramsay}, {Cannizzo}, {Howell}, {Wood},
  {Still}, {Barclay}  \& {Smale}}{{Ramsay} et~al.}{2012}]{Ramsay_PhaseJump}
{Ramsay} G.,  {Cannizzo} J.~K.,  {Howell} S.~B.,  {Wood} M.~A.,  {Still} M.,
  {Barclay} T.,   {Smale} A.,  2012, \mn@doi [\mnras]
  {10.1111/j.1365-2966.2012.21657.x}, \href
  {http://adsabs.harvard.edu/abs/2012MNRAS.425.1479R} {425, 1479}

\bibitem[\protect\citeauthoryear{{Ramsay}, {Wood}, {Cannizzo}, {Howell}  \&
  {Smale}}{{Ramsay} et~al.}{2017}]{Ramsay_729}
{Ramsay} G.,  {Wood} M.~A.,  {Cannizzo} J.~K.,  {Howell} S.~B.,   {Smale} A.,
  2017, \mn@doi [\mnras] {10.1093/mnras/stx859}, \href
  {https://ui.adsabs.harvard.edu/abs/2017MNRAS.469..950R} {469, 950}

\bibitem[\protect\citeauthoryear{{Rappaport}, {Joss}  \& {Webbink}}{{Rappaport}
  et~al.}{1982}]{Rappaport_MB0}
{Rappaport} S.,  {Joss} P.~C.,   {Webbink} R.~F.,  1982, \mn@doi [\apj]
  {10.1086/159772}, \href
  {https://ui.adsabs.harvard.edu/abs/1982ApJ...254..616R} {254, 616}

\bibitem[\protect\citeauthoryear{{Rappaport}, {Verbunt}  \& {Joss}}{{Rappaport}
  et~al.}{1983}]{Rappaport_MB}
{Rappaport} S.,  {Verbunt} F.,   {Joss} P.~C.,  1983, \mn@doi [\apj]
  {10.1086/161569}, \href
  {https://ui.adsabs.harvard.edu/abs/1983ApJ...275..713R} {275, 713}

\bibitem[\protect\citeauthoryear{{Ricker} et~al.,}{{Ricker}
  et~al.}{2009}]{Ricker_TESS}
{Ricker} G.~R.,  et~al., 2009, in American Astronomical Society Meeting
  Abstracts \#213. p.~193

\bibitem[\protect\citeauthoryear{{Rutten}, {Kuulkers}, {Vogt}  \& {van
  Paradijs}}{{Rutten} et~al.}{1992}]{Rutten_OYCar}
{Rutten} R.~G.~M.,  {Kuulkers} E.,  {Vogt} N.,   {van Paradijs} J.,  1992,
  \aap, \href {https://ui.adsabs.harvard.edu/abs/1992A&A...265..159R} {265,
  159}

\bibitem[\protect\citeauthoryear{{Scargle}}{{Scargle}}{1982}]{Scargle_LombScargle}
{Scargle} J.~D.,  1982, \mn@doi [\apj] {10.1086/160554}, \href
  {http://adsabs.harvard.edu/abs/1982ApJ...263..835S} {263, 835}

\bibitem[\protect\citeauthoryear{{Scaringi}, {Groot}  \& {Still}}{{Scaringi}
  et~al.}{2013}]{Scaringi_KIS}
{Scaringi} S.,  {Groot} P.~J.,   {Still} M.,  2013, \mn@doi [\mnras]
  {10.1093/mnrasl/slt099}, \href
  {http://adsabs.harvard.edu/abs/2013MNRAS.435L..68S} {435, L68}

\bibitem[\protect\citeauthoryear{{Shakura} \& {Sunyaev}}{{Shakura} \&
  {Sunyaev}}{1973}]{Shakura_Disk}
{Shakura} N.~I.,  {Sunyaev} R.~A.,  1973, \aap, \href
  {http://adsabs.harvard.edu/abs/1973A%26A....24..337S} {24, 337}

\bibitem[\protect\citeauthoryear{{Sklyanov} et~al.,}{{Sklyanov}
  et~al.}{2018}]{Sklyanov_NYSer}
{Sklyanov} A.~S.,  et~al., 2018, \mn@doi [Astrophysics]
  {10.1007/s10511-018-9516-y}, \href
  {https://ui.adsabs.harvard.edu/abs/2018Ap.....61...64S} {61, 64}

\bibitem[\protect\citeauthoryear{{Smak}}{{Smak}}{1971}]{Smak_UGem}
{Smak} J.,  1971, \actaa, \href
  {https://ui.adsabs.harvard.edu/abs/1971AcA....21...15S} {21, 15}

\bibitem[\protect\citeauthoryear{{Spruit} \& {Ritter}}{{Spruit} \&
  {Ritter}}{1983}]{Spruit_Gap}
{Spruit} H.~C.,  {Ritter} H.,  1983, \aap, \href
  {https://ui.adsabs.harvard.edu/abs/1983A&A...124..267S} {124, 267}

\bibitem[\protect\citeauthoryear{{Webb} et~al.,}{{Webb}
  et~al.}{1999}]{Webb_InsideOut}
{Webb} N.~A.,  et~al., 1999, \mn@doi [\mnras]
  {10.1046/j.1365-8711.1999.03002.x}, \href
  {https://ui.adsabs.harvard.edu/abs/1999MNRAS.310..407W} {310, 407}

\makeatother
\end{thebibliography}

%\begin{thebibliography}{99}
%\end{thebibliography}

%%%%%%%%%%%%%%%%%%%%%%%%%%%%%%%%%%%%%%%%%%%%%%%%%%

%%%%%%%%%%%%%%%%% APPENDICES %%%%%%%%%%%%%%%%%%%%%

%%%%%%%%%%%%%%%%%%%%%%%%%%%%%%%%%%%%%%%%%%%%%%%%%%

% Don't change these lines
\bsp	% typesetting comment
\label{lastpage}
\end{document}